\documentclass[reprint,aps,ams,amsmath,prl,twocolumn,superscriptaddress,longbibliography]{revtex4-1}
\usepackage{graphics}
\usepackage{graphicx}
\usepackage{epsfig}
\usepackage{amsmath}
\usepackage{amssymb}
\usepackage{amsfonts}
\usepackage{bm}
\usepackage{color}
\usepackage{bbm}
\usepackage{hyperref}
\usepackage[normalem]{ulem}

\renewcommand{\vec}[1]{{\bf #1}}

\usepackage{graphics}
\usepackage{subcaption}
\usepackage{xcolor}
\usepackage{bbm}
\usepackage[normalem]{ulem}
\usepackage{amssymb}
\usepackage{array}
\usepackage{amsmath}
\usepackage{graphicx}\usepackage{graphics}
\usepackage{dcolumn}
\usepackage{bm}\usepackage{varioref}
\begin{document}

\title{Giant nonreciprocity of current-voltage characteristics of noncentrosymmetric supercondctor-normal metal-superconductor junctions }

%\title{nonreciprocal current-voltage characteristics in superconductor-normal metal-superconductor junctions}
\author{T.  Liu}
\affiliation{Department of Physics,  University of Washington,  Seattle,  WA 98195,  USA}
\author{M.  Smith}
\affiliation{Materials Science Division, Argonne National Laboratory, Lemont, Illinois 60439, USA}
\author{A. V.  Andreev}
\affiliation{Department of Physics,  University of Washington,  Seattle,  WA 98195,  USA}
\author{B. Z.  Spivak}
\affiliation{Department of Physics,  University of Washington,  Seattle,  WA 98195,  USA}

%\address{Department of Physics, University of Washington, Seattle, WA 98195,  USA}

\date{\today}

\begin{abstract}
We develop a theory of   nonreciprocal current-voltage (I-U) characteristics in noncentrosymmetric superconductor-normal metal-superconductor junctions. We show that at small voltages the nonreciprocal features of the  I-U characteristics can be expressed entirely in terms of the dependence of the nonreciprocal part of the quasiparticle density of states in the normal metal part of the junction on the order parameter phase difference $\chi$  across the junction.  The amplitude of the nonreciprocity in this regime is proportional to the inelastic quasiparticle relaxation time $\tau_{in}$, and can be much larger than that in  normal materials, where it is proportional to the elastic relaxation time $\tau_{el}$.   At low bias the I-U characteristics possess additional symmetry, not present in normal conductors; they remain invariant under simultaneous reversal of current, voltage and the magnetic field.   
\end{abstract}

\maketitle

Due to Onsager's principle, the linear two-terminal conductance of time-reversal invariant systems,  $G_{l}(\bf{H})$, must be an even function of the magnetic field ${\bf H}$ \cite{Onsager,Onsager1} 
\begin{equation}
\label{eq:onsager}
G_{l}({\bf H})= G_{l}(-{\bf H}).
\end{equation}
Beyond the linear in $U$ regime,  the current-voltage ($I-U$) characteristics  in noncentrosymmetric systems can be nonreciprocal:
$J(U, {\bf H})\neq J(U,-{\bf H})$.

In  noncentrosymmetric normal conductors this phenomenon  has been investigated in several articles,  see for example Refs.~\cite{Nagaosa1,IvchenkoSpivak,Krstic,Cobden,Ideue,Rikken,Rikken1,Loss}.   
 In this case, at small ${\bf H}$ and $U$, the degree of nonreciprocity 
is proportional to the elastic relaxation time $\tau_{el}$,
\begin{equation}\label{nonrec}
\delta J= J({\bf H}) -J({\bf -H})\sim
 \eta \tau_{el} U^{2} {\bf H}.
\end{equation}
Here $\eta$ is a material dependent parameter.

 The shapes of the current-voltage (I-U)  characteristics in superconductor-normal metal-superconductor (SNS) junctions are more complicated than those of normal metals. 
The reciprocal part of the I-U characteristics of SNS junctions have been studied in many articles,  see for example Refs.~\cite{LarkinOvchinnikov1,Tinkham,Volk,Averin1, Liu}.  It was shown in Ref.~\cite{Liu} that the shape of I-U characteristics at small and large bias are qualitatively different,  as illustrated in  Figs.~\ref{fig2} and \ref{fig3}.  At relatively large bias, the I-U characteristics are controlled by the elastic mean free time $\tau_{el}$. As a result, the degree of nonreciprocity  in this regime is expected to be of order of that of the normal part of the junction, which is relatively small and featureless.  These parts of the I-U characteristics are shown in Figs.~\ref{fig2},\ref{fig3} by dashed black lines. At small bias the I-U characteristics  are controlled by the long quasiparticle inelastic relaxation time $\tau_{in}\gg \tau_{el}$,  which is typically much longer than the elastic relaxation time. Therefore, the nonlinear   conductance of the junctions   turns out to be much larger than the normal state conductance.  These parts of the I-U characteristics  are shown in Figs.~\ref{fig2} and \ref{fig3} by green and blue lines.

 There is extended literature about the nonreciprocity of supercurrent 
 in superconductors in general, and the nonreciprocity of the critical current of Josephson junctions in particular (see for example Refs.~\cite{Hasan,Ando,Fu1,Lotfizadeh,Efetov,Edelstein,Baumgartner,Baumartner1,Strunk,Turini,Lin,Nagaosa1,Fu2,LiyandaGeller,Zutich,Ilie,Li}, and references herein). On the other hand, the nonreciprocity of the dissipative part of nonlinear  I-U characteristics has attracted much less attention. The goal of this article is to develop a theory of nonreciprocity of $I-U$ characteristics of noncentrosymmetric SNS junctions in this  regime.
 We will  show that at small voltages the degree of the nonreciprocity of the I-U characteristics of SNS junctions turns out to be much larger than  in normal metals.

At small voltages $eU \ll E_{T}$,  the quasi-particle spectrum, and the I-U characteristics of the  junction can be calculated in the adiabatic approximation, treating the phase difference $\chi(t)$ as a slowly varying parameter~\cite{Liu}. Here $E_{T}$ is the Thouless energy, which is inversely proportional to the characteristic time required  for quasiparticles to travel between the two superconducting banks of the junction. For simplicity we assume that the transmission coefficient of superconductor-normal metal boundary is of order one.   In this case the charge transport through SNS junction can be expressed entirely in terms of the dependence of the density of states $\nu(\epsilon, \chi)$ on  $\chi$ and the quasiparticle energy $\epsilon$.

In the presence of the voltage,  the phase difference across the junction $\chi(t)$  evolves in time according to the Josephson relation,
\begin{equation}\label{eq:josephson}
\dot{\chi}\equiv \frac{d\chi}{dt}  = 2eU(t).
\end{equation}
Due to Andreev reflection at the normal metal-superconductor boundaries the low energy quasiparticles ($\epsilon < \Delta$)  are trapped inside the normal region, and the spectrum of these quasiparticles depends on the phase difference $\chi$.   At nonzero temperature $T$, the quasiparticles occupying these levels move in energy space together with the levels. This motion creates a nonequilibrium quasiparticle distribution, which relaxes via inelastic scattering producing a dissipative contribution to the current. 
The physical mechanism of this contribution  is similar to the Debye mechanism of microwave absorption in gases \cite{Debye}, Mandelstam-Leontovich mechanism of the second viscosity in liquids \cite{Landau}, the Pollak-Geballe mechanism of microwave absorption in the hopping conductivity regime \cite{Pollak}, and the mechanism of low frequency microwave absorption in superconductors \cite{Smith1, Smith2, Mike}.

A quantitative description of the current in the low bias regime can be obtained as follows (see for example~\cite{Liu} and references therein). 
In the adiabatic approximation, contributions to the current related to transitions between energy levels can be neglected, and the  current may be expressed as
\begin{align}\label{EQ:CURRENTEUL}
J(t ) = &   -2e \int_{0}^{\infty} d \epsilon \nu \big(\epsilon, \chi(t) \big)V_{\nu}\big( \epsilon, \chi(t) \big)  [ n(\epsilon, t )  - n_F \big(\epsilon \big) ]    \nonumber \\
&  +  J_{s} \big( \chi(t) \big).
\end{align}
Here $\nu(\epsilon, \chi)$ is the total density of states which includes spin degree of freedom, $n(\epsilon,t)$ is the nonequilibrium occupancy of quasiparticle levels with energy $\epsilon$, 
$n_F(\epsilon)$ is the Fermi distribution function at temperature $T$ (which we will assume to be much smaller than the superconducting gap  in the banks of the junction, $T \ll \Delta$),  and
$V_\nu \big(\epsilon,\chi(t) \big) $ is the sensitivity of quasiparticle energy levels to changes in $\chi$. The latter can be expressed in terms of the density of states as,
\begin{equation}\label{eq:level_sensitivity}
V_{\nu} \big(\epsilon,\chi \big) \equiv - \frac{1}{\nu (\epsilon, \chi)}  \int_{0}^{\epsilon} d \tilde{\epsilon}  \frac{\partial \nu (\tilde{\epsilon}, \chi )}{\partial \chi}.
\end{equation}
Finally, $J_{s}(\chi)$ is the supercurrent can be written as  
 \begin{equation}\label{supercurrent}
J_s(\chi) = -2e \int^\infty_0 d\epsilon \tanh\left(\frac{\epsilon}{2T}\right) \nu(\epsilon,\chi ) V_\nu(\epsilon,\chi ), 
\end{equation}
(see for example Ref.~\cite{Binn}).

The time evolution of the distribution function is described by the kinetic equation,
\begin{equation}\label{EQ:N_DOT}
 \partial_{t} n (\epsilon, t )+ \dot{\chi}\cdot  V_{\nu} (\epsilon, \chi(t))\,  \partial_\epsilon  n(\epsilon, t ) =\frac{n_F(\epsilon) - n(\epsilon,t )}{\tau_{in}},
\end{equation}
where $\tau_{in}$ is the inelastic relaxation time. Equation \eqref{EQ:N_DOT} can be derived both phenomenologically and  more rigorously using Green's functions~\cite{Liu}.

Equations \eqref{eq:josephson}-\eqref{EQ:N_DOT} together with the expression for the density of states  provide a complete description of charge transport through SNS junctions provided the phase evolution rate is sufficiently slow.  Thus, in this regime the I-U characteristics of the junctions are determined by $\epsilon$ and $\chi$  dependence of the density of states. 
%They were studied  in Ref.~\cite{Liu}  and have the shape shown, respectively, in Figs.~\ref{fig2} and \ref{fig3}. 

 In general,  both the current and voltage across junctions exhibit oscillations in time. 
 We will be interested in the shape of the I-U characteristics averaged over the period of oscillations,  and will indicate the time-averaged quantities by overline,  e.g.  $\bar{J}$ , and $\bar{U}$.

 In nonmagnetic systems the density of states  is invariant under the change of sign of both $\chi$ and $\bm{H}$, $\nu(\epsilon, \chi, {\bf H}) = \nu(\epsilon, -\chi, -{\bf H})$, which implies 
\begin{equation}
\label{eq:TRS_condition}
V_\nu (\epsilon, \chi, {\bf H}) = -V_\nu(\epsilon, -\chi, -{\bf H})
\end{equation}
for the level sensitivity.  

It follows from  Eqs.~\eqref{eq:josephson}-\eqref{eq:TRS_condition}  that in the low bias regime considered here the I-U characteristics in SNS junctions  possesses a symmetry, which is not present in devices based on normal conductors (see  e.g.  Ref.~\cite{Glazman}), and is independent of the device geometry. Namely,  it is invariant under the simultaneous reversal of the magnetic field, current and voltage, 
\begin{align}
\label{eq:nonreciprocity_symmetry}
\bar{J}(- \bar{U}, - {\bf H}) = - \bar{J}( \bar{U},  {\bf H}).
\end{align}
Thus, at low bias one can equivalently define nonreciprocity as part of the I-U characteristic, which is odd under the reversal of current and voltage for a fixed magnetic field,  
$\delta \bar{J}=\bar{J}(\bar{U},{\bf H})-J(\bar{U},- {\bf H})=\bar{J}(\bar{U},{\bf H})+J(-\bar{U}, {\bf H})$.  

The nonreciprocity of the I-U characteristics in the low bias regime described by Eqs.~\eqref{eq:josephson}-\eqref{eq:TRS_condition}  arises from the odd in $\bf {H}$ dependence of the density of states, $ \delta \nu(\epsilon, \chi, {\bf H}) =\nu(\epsilon, \chi, {\bf H})-\nu(\epsilon, \chi, -{\bf H})$.
The latter depends on the orientation of the ${\bf H}$ relative to the current. 
It is important to note that in the special case where the influence of ${\bf H}$ on $ \nu(\epsilon, \chi, {\bf H})$  reduces a constant phase shift  $\phi({\bf H})$,  
\begin{equation}\label{nuH}
\nu(\epsilon,\chi, {\bf H})=\nu_{0}(\epsilon, \chi + \phi({\bf H})) ,
 \end{equation}
 the I-U characteristics  remain reciprocal. 
%{\color{blue}The reason for this, as we will show below, neither the average current in the fixed voltage case nor the average voltage in the fixed current case are affected by the phase shift }
The reason is that neither the average voltage $\bar{U}$ nor the average current $\bar{J}$ are affected by the phase shift.  
Furthermore, it follows from Eq.~\eqref{supercurrent}  that in this case the critical current of the junction is also reciprocal \cite{Hasan}.
 We note however,  that if different junctions obeying Eq.~\eqref{nuH} are connected in parallel~\cite{K.C.Fong,Fominov}, the critical current and resistance at current bias are nonreciprocal.

%%%%%%%%%%%%%%%%%%%%%%%%%%%%%%%%%%%%%%%%%
\begin{figure}[h!]
\centering
\includegraphics[width=0.49\textwidth]{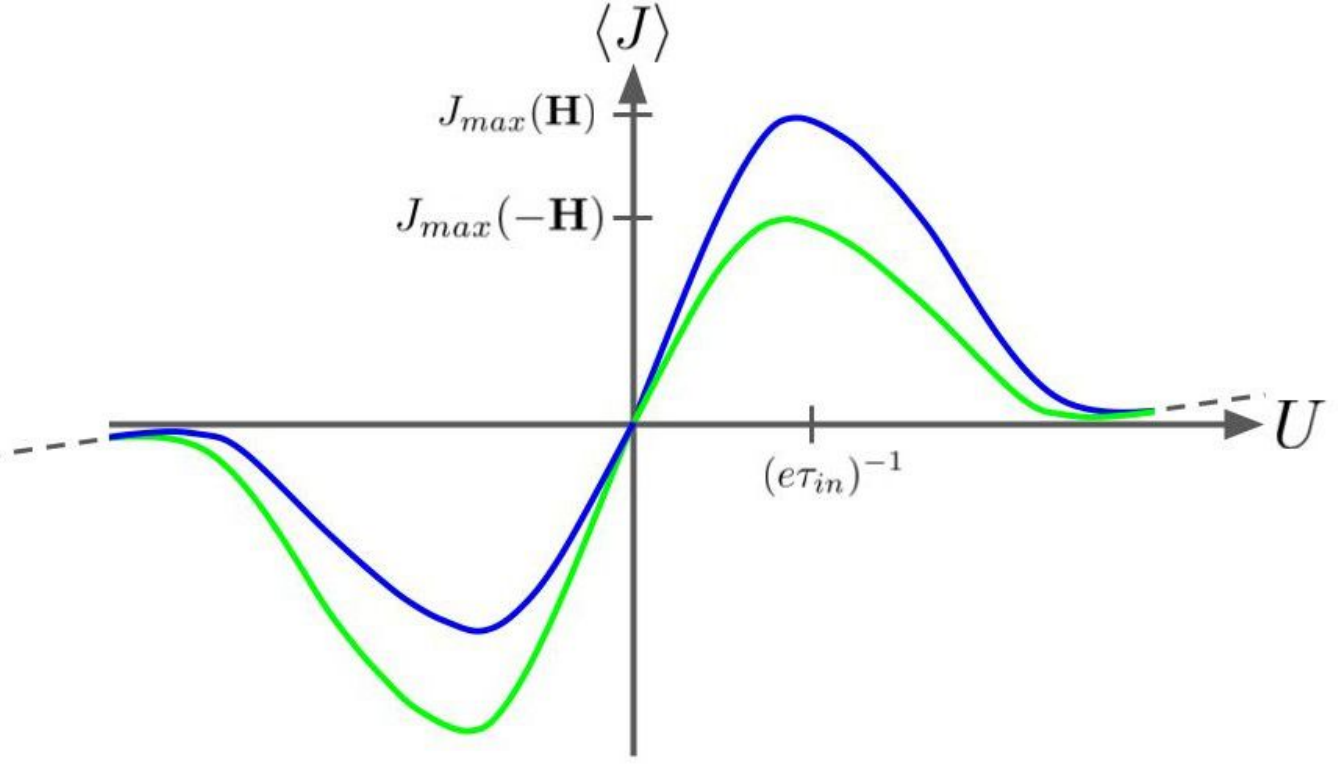}
\caption{The I-U characteristics of a nonreciprocal SNS junction at low voltage bias for opposite signs of the magnetic field are sketched in blue and green. The dashed lines correspond to the high voltage regime.
%\emph{Inset:} geometry of planar SNS junction; inversion symmetry is broken by out of plane vector $\hat{\bf n} \parallel \hat{z}$, the junction is aligned along the $ \hat{\bf x}$ 
%direction, and the magnetic field ${\bf H}$ is directed in the $\hat{\bf  y}$ direction.
}
\label{fig2}
\end{figure}
%%%%%%%%%%%%%%%%%%%%%%%%%%%%%%%%%%%%%%%%%%%%

The shapes of I-U characteristics of SNS junctions depends on the external circuits, we will consider below two limiting cases of voltage and current biased junctions.

\emph{Voltage bias:} We begin the consideration of nonreciprocity with the voltage bias setup.   The I-U characteristic in this case has an $N$ shape ~\cite{Liu}, sketched  in Fig.~\ref{fig2}.  At  low bias,  where the rate of phase evolution obeys the inequality $\dot{\chi}(t)\tau_{in} \ll 1$ the solution to Eq.~\eqref{EQ:N_DOT} may be expressed as a series in the small parameter $\dot{\chi}(t)\tau_{in}$. To  second order  we obtain for the instantaneous current 
\begin{align}\label{localcurrent}
J(t,T,{\bf H}) = & \,   J_{s} \big(\chi(t), T,{\bf H}\big)+ g_1 \big(\chi(t),{\bf H} \big) \big(\dot \chi(t) \tau_{in} \big) \nonumber \\
&+ g_2 \big(\chi(t) ,{\bf H}\big) \big(\dot \chi(t) \tau_{in} \big) ^2 ,
\end{align}
where 
\begin{subequations}
\label{eq:g12}
\begin{align}
\label{g11}
g_{1}(\chi, {\bf H}) = &\, 
-2e \int_{0}^{\infty}  d \epsilon\partial_\epsilon n_F(\epsilon) \nu(\epsilon, \chi, {\bf H}) V^2_{\nu}(\epsilon, \chi, {\bf H}), \\
\label{g22}
g_2(\chi,{\bf H}) = &\,  e \int^\infty_0 d\epsilon
\bigg\{\partial_\epsilon n_F(\epsilon) \partial_\chi \left[ \nu(\epsilon,\chi, {\bf H}) V^2_\nu(\epsilon,\chi, {\bf H}) \right]\nonumber \\
&+  \partial_\epsilon^2 n_F(\epsilon) \nu(\epsilon,\chi,{\bf H}) V_\nu^3(\epsilon,\chi,{\bf H}) \bigg\}.
\end{align}
\end{subequations}
At $\dot{\chi}(t) = 2 e U  =\mathrm{const}$, the average current can be obtained by averaging Eq.~\eqref{localcurrent} over the phase $\chi$.  The supercurrent, 
averages to zero.  The linear conductance arises from the second term on the RHS,  and has the form 
\begin{align}
\label{eq:G_l_g1}
G_l({\bf H}) = 2 e \tau_{in} \langle g_1(\chi,{\bf H})\rangle,
\end{align}
where $\langle \ldots \rangle$ denotes averaging over the phase $\chi$.  Using Eqs.~  \eqref{eq:TRS_condition}  and \eqref{g11} it is easy to see   that the linear conductance is an even function of ${\bf H}$ in accordance with the Onsager symmetry principle.

The nonreciprocal part of the \emph{dc} current $\delta \bar{J}(U, {\bf H}) = \bar{J}(U,{\bf H}) - \bar{J}(U,- {\bf H})$ arises from the third term on the RHS of Eq.~\eqref{localcurrent}, and has the form 
\begin{align}\label{avgcurrent}
\delta \bar{J} (U, {\bf H}) & = 2 \, 
\langle g_2(\chi, {\bf H})  \rangle  
\big(2eU \tau_{in}\big) ^2.
\end{align}
It follows from  Eq.~\eqref{eq:TRS_condition} that
\[
\langle g_2(\chi, -{\bf H})  \rangle   = - \langle g_2(\chi, {\bf H})  \rangle.
\]
 is an odd function of ${\bf H}$, in agreement with Eq.~\eqref{eq:nonreciprocity_symmetry}.
 It is interesting that the linear conductance in Eq.~\eqref{eq:G_l_g1} is proportional to $\tau_{in}$, while the nonreciprocal current in  Eq.~\eqref{avgcurrent} is proportional to $\tau^2_{in}$.

In the voltage bias setup it is possible to obtain closed-form expressions for the I-U characteristic for arbitrary values of $2e U \tau_{in}$.  The results are presented in the Appendix.  Here we summarize their main features.
The magnitude of the current reaches its maximum at $eU\tau_{in}\sim 1$, and the values of the maximum current 
$J_{max}({\bf H})$
and its nonreciprocity $\delta J_{max}({\bf H}) = J_{max}({\bf H}) - J_{max}(-{\bf H})$ may be estimated as 
\begin{eqnarray} \label{eq:J_max}
J_{max}({\bf H})  \sim  \langle g_{1}(\chi, {\bf H}) \rangle , \quad 
\delta J_{max}({\bf H})  \sim  \langle g_2(\chi, {\bf H}) \rangle.
\end{eqnarray}

We note that at sufficiently large temperatures the ciritical current becomes exponentially small in $T$, while according to Eqs. ~\eqref{eq:J_max} and \eqref{g11},  the maximal current $J_{max}$ does not have such strong temperature dependence. Therefore,
$J_{max}$ can be much larger than $J_{c}$ \cite{Liu}.

At relatively large voltages $eU\tau_{in}\gg 1$ the solutions of Eqs.~\eqref{eq:josephson} -\eqref{EQ:N_DOT} yield an expression for the current  which decreases with voltage,
\begin{equation}\label{avgcurrent2}
\bar{J} (U, {\bf H}) \sim \frac{C_{1}({\bf H})}{eU\tau_{in}} +  \frac{C_2({\bf H})}{(eU\tau_{in})^2}.
\end{equation}
Here to within a factor of order unity   $C_1 \sim  \langle g_1(\chi, {\bf H} ) \rangle$ and $C_2  \sim  \langle g_2(\chi, {\bf H}) \rangle$.   

Eventually,  at large voltages the  I-U characteristics are controlled by the contribution to the current which is beyond the adiabatic approximation and is of order $G_{N}U$.  Here $G_N  $ is the conductance of the normal metal part of the junction.  As the voltage approaches  this  regime, the current reaches a minimum at $U=U_{min}$.
At this point, the contribution to the conductance proportional to $\tau_{in}$ and $\tau_{el}$ are of the same order. The position of the minimum is approximately given by,
\begin{equation}
\label{eq:U_min}
U_{min} ({\bf H}) \sim  \frac{1}{\tau_{in}} \sqrt{\frac{G_l({\bf H}) }{G_N}}
+\frac{ \langle g_2({\bf H}) \rangle }{\langle g_{1}({\bf H})) \rangle} {\frac{G_l({\bf H}) }{G_N}}  .
\end{equation}
At $U\gg U_{min}$ the I-U characteristics of the system   (including  its nonreciprocity)   is roughly the same as that of the normal part of the SNS junction.

Equations \eqref{avgcurrent}-\eqref{eq:U_min} show that  the general key features of nonreciprocity of the I-U characteristics of voltage-biased junctions are determined by   the parameters  $\langle g_1 (\chi,{\bf H})\rangle$ and $\langle g_{2} (\chi, {\bf H}, )\rangle$.  Note that the first term  on the HRS of Eq.~\eqref{g22} is a total derivative and averages to zero.   
In many physical situations 
 the relevant quasiparticle energies 
 are of order of the Thouless energy $E_{T}$.  
In this  case, for  $T\gg E_{T}$  Eqs.~\eqref{eq:g12} 
simplify to 
\begin{subequations}
\label{eq:g12_high_T}
\begin{align}
\label{g11_high_T}
\langle g_{1}(\chi, {\bf H}) \rangle  = & \, 
\frac{e}{2T}  \int_{0}^{\infty}  d \epsilon   \langle\nu(\epsilon, \chi, {\bf H}) V^2_{\nu}(\epsilon, \chi, {\bf H})\rangle , \\
\label{g22_high_T}
\langle g_2(\chi,{\bf H})\rangle  = &\,   \frac{e}{32 \pi T^3}  \int^\infty_0 d\epsilon  \, \epsilon \, \langle
 \nu(\epsilon,\chi,{\bf H}) V_\nu^3(\epsilon,\chi,{\bf H}) \rangle.
\end{align}
\end{subequations}
and we get  
\begin{equation}\label{eq:EtT}
\frac{\langle g_2(\chi, {\bf H},T)  \rangle}{\langle g_1(\chi,T)  \rangle}     =
\gamma({\bf H})\frac{E_{T}^2}{T^2},
\end{equation}
where $\gamma({\bf H})$ is an odd function of $ {\bf H} $, which  is determined by the sensitivity of the phase-dependent quasiparticle spectrum on the magnetic field.   The different temperature dependence of $\langle g_1 (\chi,{\bf H})\rangle$ and $\langle g_{2} (\chi, {\bf H} )\rangle$ in Eq.~\eqref{eq:g12_high_T} arises because
the second term in Eq.~\eqref{g22} requires going to higher order in the Sommerfeld expansion. 

%{\color{blue}
%Finally we note that for a density of states in the form of Eq.~\eqref{nuH}, the coefficients $\langle g_{1,2}(\chi, {\bf H}) \rangle$ are equal to their values at ${\bf H} = 0$. This is because $g_{1,2}(\chi,{\bf H})$ are periodic functions of $\chi$ and their average values are unaffected by a constant phase shift. As a result, $\langle g_{2}(\chi, {\bf H}) \rangle$ vanishes in this case and there is no non-reciprocity.
%}

\emph{Current bias:}  Let us now turn to the consideration of nonreciprocity   in the current-bias setup.  The shape of the I-U characteristic in this  case~\cite{Liu} is illustrated in  Fig.~\ref{fig3}. 
The nonreciprocity of the critical current $J_{c}$ of SNS junctions has been studied in several articles \cite{Hasan,Baumgartner,Ando,Fu1,Lotfizadeh,Efetov,Edelstein,Baumartner1,Turini,Lin,Nagaosa1,Fu2,LiyandaGeller,Zutich, Ilie}. Here we study the I-U characteristics $\bar{U}(J)$ at currents larger than the critical current.

Similarly to the voltage-biased case, the dissipative part of the I-U characteristics can be divided into two parts. At $\bar{U}>1/\tau_{in}$ and $\bar{U}< 1/\tau_{in}$ it is controlled by the elastic and inelastic relaxation times respectively.  The  description of the I-U characteristics in the adiabatic approximation is valid  at relatively small average voltages $\bar{U}< 1/\tau_{in}$  when $J<J_{jump}$ \cite{Liu}.
At higher currents, $J>J_{jump}$,  the voltage increases dramatically. In this regime the dissipation is determined by the elastic relaxation time, and the shape of this part of the I-U characteristics is,  roughly speaking, the same as in the normal metal. Therefore, similar to the case of current-biased junctions, the degree of nonreciprocity in this region is expected to be relatively small.

\begin{figure}[h]
\includegraphics[width=0.49\textwidth]{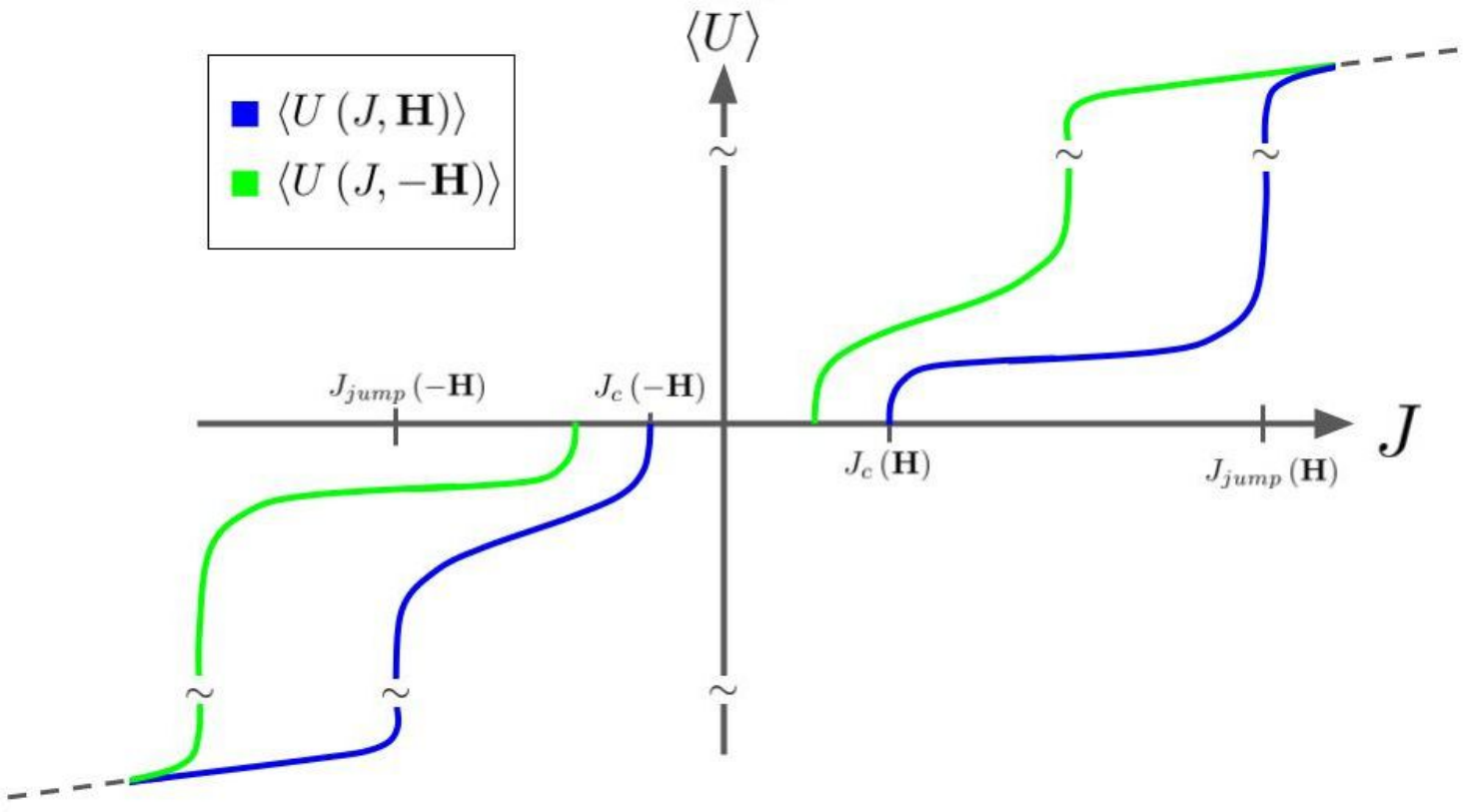}
\caption{A qualitative picture of the I-U characteristics of a nonreciprocal SNS junction at fixed current. The blue and green curves correspond to  the low voltage  regime regime of the I-U characteristics for opposite signs of magnetic field. The dashed lines correspond to the high voltage regime. }
\label{fig3}
\end{figure}

In the current bias setup  the phase difference  $\chi (t)$ increases monotonically with time, but at a varying rate. Using the Josephson relation Eq.~\eqref{eq:josephson}
we can relate the average voltage across the junction to the duration $t_p$ of the time interval  during which $\chi$ increases by   $2\pi$,
\begin{equation}\label{average_voltage}
\bar U = \frac{\pi}{e t_p}.
\end{equation}
To evaluate  $t_{p}$, we must determine the dynamics of $\chi(t)$.

When the current is close to the critical current,  $J - J_c({\bf H}) \ll J_c ({\bf H}) $, the dominant contribution to the period $t_P$ comes from the  interval of phase where $\chi(t)$ is near the phase $\chi_m({\bf H})$, at which the supercurrent reaches its maximum value   (see \cite{Liu}).  In this interval  $\dot\chi(t) \ \ll 1/\tau_{in}$,  and   Eqs.~\eqref{localcurrent}-\eqref{eq:g12} can be used, leading to  
\begin{equation}\label{tp2}
t_p \approx  \tau_{in}  \int^\pi_{-\pi}  \frac{    g_1(\chi_m({\bf H}) ,{\bf H})   d\chi }{J - J_s(\chi, {\bf H},T)}.
\end{equation}
Using the quadratic phase dependence of $J_s(\chi, {\bf H},T)$ near $\chi =\chi_m({\bf H})$,  $J_s(\chi, {\bf H},T)  \approx J_c({\bf H},T) +  \frac{(\chi - \chi_m({\bf H}))^2}{2}  \partial^2_{\chi}   \left.J_s(\chi, {\bf H},T)\right|_{\chi = \chi_m({\bf H})} $ we can express the average voltage  as 
\begin{equation}\label{avgU2}
 \bar U( J,{\bf H}) \approx A ({\bf H}) \sqrt{(J - J_c({\bf H},T))},
\end{equation}
where 
\begin{align}
\label{eq:A_H}
 A ({\bf H}) = &\, \frac{\sqrt{- \partial^2_{\chi}   J_s(\chi, {\bf H},T)|_{\chi = \chi_m({\bf H})}  }}{2^{5/2}    e \tau_{in} g_1(\chi_m({\bf H}) ,{\bf H}) }. 
\end{align}
The voltage in Eq.~\eqref{avgU2} has the standard square root dependence on the excess current $J - J_c({\bf H},T)$. The nonreciprocity in this regime is characterized by the nonreciprocity of the critical current $J_c({\bf H},T)$ and the coefficient $A({\bf H})$.  Equation ~\eqref{eq:A_H} shows that nonreciprocity of $A({\bf H})$ is determined  not only by the sensitivity of the phase-dependent supercurrent to the magnetic field but also by the dissipative parameter $g_1(\chi)$. Indeed,  the phase $\chi_{m}({\bf H})$ at which the supercurrent attains the maximal value in the direction of the bias shifts linearly with ${\bf H}$.  Since $g_1(\chi)$ does not have an extremum at $\chi_{m}({\bf H})$ the denominator in Eq.~\eqref{avgU2} contributes to nonreciprocity of $A ({\bf H})$. 

%{\color{blue}
%Here we also note that the integrand in Eq.~\eqref{tp2} is a function of the density of states, which is a periodic function of $\chi$. Thus, for for a density of states in the form of Eq.~\eqref{nuH}, $t_p$ is equal to its value at ${\bf H} =0$ and the non-reciprocity vanishes like in the fixed voltage case.}

It was shown in Ref.~\cite{Liu},  that
\begin{equation}
J_{jump}\sim J_{max} ,
\end{equation}
where $J_{max}$ was determined for the voltage bias case.  Interestingly, it follows from our equations that at $T>E_{T}$  the degree of nonreciprocity for current bias, $\delta J_{jump} =J_{jump}({\bf H})-J_{jump}(-{\bf H})$, is enhanced with respect to the voltage bias case, Eqs.~\eqref{eq:J_max},\eqref{eq:EtT}, by the factor $T^2/E_{T}^2$\begin{eqnarray}
\frac{\delta J_{jump}}{\delta J_{max}}     \sim  \left\{   
\begin{array}{cc}
1, & T \lesssim E_T,  \\
\frac{T^{2}}{E_{T}^{2}},    &  T>E_{T}.
\end{array}
\right.
\end{eqnarray}
The reason for this can be traced to the non-constant phase evolution rate $\dot{\chi}(t)$.  Because of this the leading term in the Sommerfeld expansion of $g_2(\chi, {\bf H})$ (first term in Eq.~\eqref{g22}) can no longer be written as a total derivative of some function of $\chi$, and its contribution to the average voltage does not vanish.

The results presented above assume that the low-energy quasiparticles are trapped inside the normal region of the junction by insulating boundaries and Andreev reflection from the S-N boundaries, and the only channel of quasiparticle relaxation is inelastic scattering. 
In situations where escape of quasiparticles from the normal region is possible, $\tau_{in}$ should be replaced by the  characteristic time of quasiparticle escape from the normal region.

The above consideration can also  be extended to the cases where the nonreciprocity of the I-U characteristics is associated with existence of a spontaneous magnetization ${\bf M}$ (or spontaneous valley symmetry breaking in twisted graphene  \cite{Efetov,Lin,Scammell}) in the normal region of the junction at ${\bf H}=0$. In this case the nonreciprocal part of the I-U characteristics can be expressed in terms of $\delta \nu =\nu(\epsilon, \chi, {\bf M})- \nu(\epsilon, \chi, -{\bf M})$.

To estimate the magnitude of the effect, below we apply the general results obtained above to a planar junction of length $L$ and width $L_1$(shown in Fig.~\ref{fig4}),  in which the normal region is  described by the following Hamiltonian, \color{black}
\begin{align} \label{hamiltonian}
	H &= {\bf p}^2/2m - E_{F} + \beta^{\alpha i} p^i \sigma^\alpha 
	 +  V_{imp}({\bf r}) + g \mu_0 {\bf H} \cdot {\boldsymbol \sigma} .
\end{align}
Here $E_{F}$ is the Fermi energy, $m$ is the electron mass, $\sigma_i$ are the Pauli matrices in spin space, $g$ is the g-factor, $\mu_0$ is the Bohr magneton, and $V_{imp}({\bf r})$ is the random impurity potential.
For Rashba spin-orbit coupling $\beta^{\alpha i} = \alpha_R \epsilon^{\alpha i j}\hat n^j$, where $\hat n$ is a unit polar vector, and for Dresselhaus spin-orbit coupling  $\beta^{\alpha i} = \alpha_D \delta^{\alpha i}$. 

The direction of the magnetic field  is chosen to be parallel to the film,  as depicted in Fig.~\ref{fig4}. Therefore,  it enters the Hamiltonian, Eq.~\eqref{hamiltonian},  only  via the Zeeman term. 
%{\color{red} We consider two models:   \emph{i)}  $H_{0}= {\bf p}^2/2m - E_{F}$ (a parabolic band with weak spin-orbit coupling $\beta\ll v_{F}$ ),  and \emph{ii)}; the surface of a topological insulator, where $\beta\lesssim v_{F}$ and  $H_{0}=-E_{F}$.  Here $E_{F}$ is the Fermi energy, $v_F$ is the Fermi-velocity, and $m$ is the electron mass. 
%The two models may be viewed as limiting cases of weak and strong spin-orbit  coupling.   }

Below we will focus on linear in ${\bf H}$ contribution to the nonreciprocity of the I-U characteristics. 
We consider a case of weak spin-orbit coupling  $\beta p_F \ll \tau_{el}^{-1} $ and focus on  the diffusive regime,  $L\gg \sqrt{D\tau_{so}}$. 
Here $D=v_{F}^{2}\tau_{el}/2$ is the typical value of the electron diffusion coefficient in the normal region, $\tau_{so}$ is the spin relaxation time, and  $v_F$ is the Fermi-velocity. 
We also assume that the distance between the superconductors, $L$, is much larger than the  coherence length in the superconductors,  and therefore the  order parameter $\Delta({\bf r})$ has a constant   modulus $\Delta$ in the superconducting leads, and vanishes in the normal region, see Fig.~\ref{fig4}. 

\begin{figure}[h]
\includegraphics[width=0.49\textwidth]{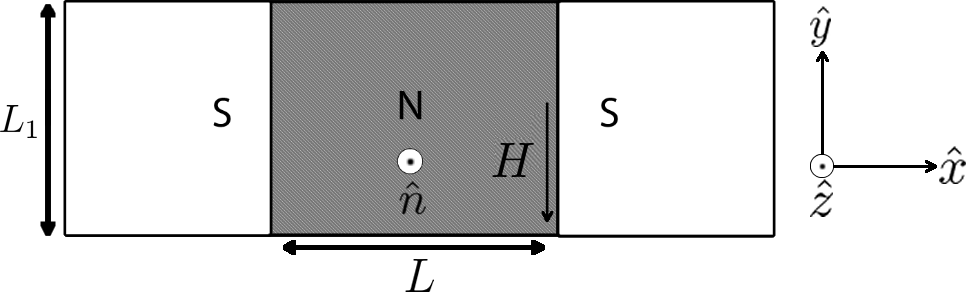}
\caption{Top down view of a planar SNS junction. The junction is aligned along the $ \hat{\bf x}$ direction, there is a parallel magnetic field ${\bf H}$ directed in the $\hat{\bf  y}$ direction, and  there is an out of plane vector $\hat{\bf n}$ pointing in the $\hat{\bf z}$ direction which breaks inversion symmetry.}
\label{fig4}
\end{figure}

After averaging over the random impurity potential,  the density of states in the SNS junction $\nu(\epsilon,\chi,{\bf H})$ is obtained by solving Usadel's equation in the presence of spin-orbit coupling and  a magnetic field  \cite{Loss, tokaltly,Ilie}.  Here we  present the main results leaving the details of the calculations to Appendix  B .

The main feature of the density of states of a diffusive SNS junction is the existence of a mini-gap at $\chi=0$  of order  $E_{T}\sim D/L^{2}$ \cite{Atland,Melsen,Frahm,ZhouSpivak} (for simplicity we restrict ourselves to the case where the S-N boundaries of the junction are transparent).  The density of states $\nu(\epsilon, \chi, {\bf H})$  exhibits a significant $\chi$-dependence only for energies of the order of the mini-gap.
 This means that the level sensitivity, Eq.~\eqref{eq:level_sensitivity},  is peaked in the energy interval $\epsilon \gtrsim E_{T}$.  

We show in Appendix B that if
 the diffusion coefficient $D(x)$ and the strength of spin-orbit coupling $\beta^{\alpha x}(x)$  depend only on the $x$ coordinate,
the density of states at ${\bf H}\neq 0$  can be written in the form Eq.~\eqref{nuH} with
\begin{equation}\label{shift}
\phi ({\bf H}) = \int^{L/2}_{-L/2} dx \frac{2\tau_{so} g \mu_0 \beta^{\alpha x}(x)  {\bf H} ^\alpha }{D(x)}.
\end{equation}
 Therefore,  in this idealized 1D model the I-U characteristics  are reciprocal.  
However, in the general case where $D({\bf r})$ and $\beta({\bf r})  $ are functions of the
 two coordinates, $x$ and $y$, or the shape of the normal metal part of the junction is not rectangular, the I-U characteristics of the junction are non-reciprocal. 

Below we estimate the degree of non-reciprocity of the I-U characteristics in the case where $L\lesssim L_{1}$, the amplitude of fluctuations the diffusion coefficient in the y-direction is of order $\delta D$, and the correlation length of such fluctuations is of order $L_{1}$. 
In this case we get the following estimates for $\langle g_1 \rangle$ and $\langle g_2 \rangle$,
%\begin{subequations}
\begin{align}
\label{g1diffusive}
\langle g_1 \rangle  & \sim
e \nu_N \frac{E_T^3}{T}, \quad 
\langle g_2 \rangle  \sim \langle g_1 \rangle
 \frac{\beta (g \mu_0 H) \tau_{so} E_T }{L  T^2 } 
\bigg( \frac{\delta D}{D} \bigg)^2 ,
\end{align}
%\end{subequations}
where  $\tau_{so}^{-1} = 4 p_F^2 \beta^2 \tau_{el}$.
For the nonreciprocal part $\delta A\equiv[ A({\bf H})  - A(-{\bf H}) ]$ of the coefficient $A({\bf H})$  in Eq.~\eqref{eq:A_H} we find
\begin{equation}\label{deltaA}
\begin{split}
\delta  A \sim \frac{\sqrt{J_c(0,T)} }{e\tau_{in} \langle g_1 \rangle}
\frac{\tau_{so} \beta ( g \mu_0 H) }{L E_T}  \bigg( \frac{\delta D}{D} \bigg)^2 .
\end{split}
\end{equation}
The parameters  in Eqs.~\eqref{g1diffusive}--\eqref{deltaA}   together with Eqs.~\eqref{avgcurrent} and \eqref{avgU2}  characterize  the degree of nonreciprocity of I-U characteristics of  SNS junctions with weak spin-orbit coupling in both the voltage and current bias cases.

{We note that transport in diffusive junctions in which the normal region is formed by a surface of a topological insulator can also be analysed using Usadel's equation. This is done in Appendix B. The magnitude of nonreciprocity in this case is obtained by setting  in Eqs.~\eqref{g1diffusive} and \eqref{deltaA}    $\tau_{so} \to \tau_{el}$ and   $\beta \to v$, where $v$ is the velocity of the relativistic dispersion.  
}

%{\color{red} We note that our estimates can be extended to superconductor-topological insulator-superconductor  junctions, where the spin-orbit coupling is strong $\beta \lesssim v_F$. In this case, the spin relaxation time $\tau_{so}$ which appears in Eqs.~\eqref{g1diffusive} and \eqref{deltaA} should be replaced with the elastic relaxation time $\tau_{el}$  (see Appendix B for details).}

We conclude by summarizing our main results. We have shown that at low bias the nonreciprocal features of the  I-U characteristics in SNS junctions can be expressed in terms of the nonreciprocal part of the density of states, 
$\delta \nu =\nu(\epsilon, \chi, {\bf H})- \nu(\epsilon, \chi, -{\bf H})$.  This leads to an a symmetry not present for normal conductors; the I-U characteristics remain invariant under a simultaneous reversal of curr
ent, voltage and magnetic field. At low bias all features of I-U characteristic are controlled by the inelastic relaxation time.  This leads to a much stronger nonreciprocity in comparison to normal conductors.  Although the maximal current $J_{max}$ in the voltage bias setup and the current $J_{jump}$ at which the voltage rapidly rises in the current bias setup are of the same order, their nonreciprocity turn out to be parametrically different at $T>E_T$.

This work of T. L.  and A. A. was supported by the US National Science Foundation through the MRSEC Grant No. DMR-1719797,  the Thouless Institute for Quantum Matter (TIQM),  and the College of Arts and Sciences at the University of Washington.

\begin{widetext}

\title{Supplemental Material: Giant nonreciprocity of current-voltage characteristics of noncentrosymmetric supercondctor-normal metal-superconductor junctions }
\maketitle

\appendix 

\section{Appendix A: Derivation of general expression for the current of voltage-biased junctions}
\label{fixedvoltage}
\setcounter{equation}{0}
\renewcommand{\theequation}{A\arabic{equation}}

To obtain a general expression for the current in the fixed voltage case, it is useful introduce the variable N, defined as the density of states integrated up to energy $\epsilon$,
\begin{equation}\label{EqN}
N = \int^{\epsilon(N,\chi(t), {\bf H} )}_0 d \tilde \epsilon \nu(\tilde \epsilon,\chi(t),{\bf H} ),
\end{equation}
and make a change of variables from $\epsilon$ to $N$ in Eqs.~(4)--(7). 
Following this change of variables, the function $\epsilon \big(N, \chi(t), {\bf H} \big)$ is to be interpreted as the energy of the Nth level counting from $\epsilon = 0$. This change of variables is similar to changing from Euler to Lagrange variables in hydrodynamics. 
 Equation (7)  written with the variables $\big(N, \chi(t), {\bf H} \big)$ has the form,
\begin{equation}
\label{kinequation2}
\partial_t n(N,t) = \frac{n_F \big( \epsilon(N,\chi(t), {\bf H} )\big) - n(N, \chi(t))}{\tau_{in}}.
\end{equation}
At $t \gg \tau_{in}$ the solution to Eq.~\eqref{kinequation2} is insensitive to initial conditions, 
\begin{equation}\label{generalN}
n(N,t) = \int^\infty_0 \frac{d\tau}{\tau_{in}} e^{-\frac{\tau}{\tau_{in}} } n_F\big( \epsilon \big( N,\chi(t-\tau), {\bf H} \big)\big).
\end{equation}
Substituting it into  Eq.~(4) and averaging over the period of oscillations, we get the following expression for the average current,
\small
\begin{equation}\label{eqA:Jgeneral}
\bar J (U, {\bf H}) = \frac{e}{\pi} \int^\infty_0 dN \int^\pi_{-\pi} d\chi \partial_\chi \epsilon(N,\chi, {\bf H}) \int^\infty_0 \frac{d\phi}{\lambda} e^{-\phi / \lambda} 
n_F\big( \epsilon \big(N,\chi -\mathrm{sgn}(U) \phi , {\bf H}\big)\big).
\end{equation}
\normalsize
Here  $\lambda = |2eU\tau_{in}|$,  and we changed the integral of $\tau$ to an integral over $\phi = |2eU\tau|$.  Since we have converted the integrals over time to integrals over phase, we omit the time dependence of $\chi(t)$ for the rest of the appendix. \color{black}
%We have also used the fact that that the phase winds at a constant rate in the fixed voltage case, averaging over time is equivalent to average over $\chi$.

In the regime  $E_{T}/T\ll 1$, where the typical distance of level motion is small compared to the temperature, we can expand the Fermi function about the average energy of the Nth energy level $\bar \epsilon(N, {\bf H}) = 1/2\pi \int^{\pi}_{-\pi} d\chi \epsilon(N, \chi, {\bf H})$. Keeping terms up to second order in deviations from the average value, we substitute this expansion into Eq.~\eqref{eqA:Jgeneral} and obtain the following expression for the current,
\begin{align}
\bar J (U,{\bf H})  =&\, \frac{e}{\pi} \int^\infty_0 dN  \int^\pi_{-\pi} d\chi \partial_\chi \epsilon(N,\chi, {\bf H}) \int^\infty_0 \frac{d\phi}{\lambda} e^{-\phi / \lambda} 
\bigg(n_F\big(N, \bar \epsilon(N, {\bf H}) \big) \nonumber \\
&+\partial_\epsilon n_F\big( \bar \epsilon(N, {\bf H}) \big) 
\big(\epsilon(N, \chi - \mathrm{sgn}(U) \phi, {\bf H}) - \bar \epsilon(N, {\bf H}) \big) \nonumber \\
&+\frac{1}{2} \partial^2_\epsilon n_F\big(\bar \epsilon(N,{\bf H}) \big)
\big(\epsilon(N, \chi - \mathrm{sgn}(U)\phi, {\bf H}) - \bar \epsilon(N, {\bf H})\big)^2
\bigg ).
\end{align}
To simplify this expression, it is useful to define the Fourier expansion of $\epsilon(N,\chi)$,
\begin{equation}
\epsilon(N,\chi, {\bf H}) = \bar \epsilon(N, {\bf H}) + \sum_{k} C_k(N, {\bf H}) e^{ik\chi}.
\end{equation}
The coefficients must satisfy $C_k (N, {\bf H})  = C^*_{-k} (N, {\bf H})  $ since the LHS is real. 
Furthermore, time reversal symmetry ensures that $\epsilon(N, \chi, {\bf H}) = \epsilon(N, -\chi, -{\bf H})$. As a result, we also have $C_k(N,{\bf H}) = C_{-k} (N, -{\bf H})$.

Substituting this into the expression for the current and evaluating the integrals of $\phi$, we get
\begin{align}\label{generalJ}
\bar J (U, {\bf H} )=& (2e) \mathrm{sgn}(U) \int^\infty_0 dN \partial_\epsilon n_F\big(\bar \epsilon(N, {\bf H}) \big) \sum_{k} |C_k(N, {\bf H})|^2  \bigg( \frac{-  k^2 \lambda }{1 + (k\lambda)^2} \bigg)\nonumber \\
&-e \int^\infty_0 dN \partial^2_\epsilon n_F\big(\bar \epsilon(N, {\bf H}) \big) \sum_{k_1,k_2} C_{k_1}(N, {\bf H}) C_{k_2}(N, {\bf H}) C_{-k_1 - k_2}(N,{\bf H})  \nonumber \\
& \times 
 (k_1 + k_2)  \bigg( \frac{i + \mathrm{sgn}(U) \lambda  (k_1 + k_2) }{1 + (k_1 + k_2)^2\lambda^2} \bigg).
\end{align}
\normalsize
Let us separate the average nonlinear current into the reciprocal, $\bar J_{rec} (U, {\bf H} )$, and nonreciprocal, $\delta \bar J (U, {\bf H})$ parts,
\begin{align}
\label{eq:J_rec_nonrec}
\bar J_{\mathrm{rec}} (U, {\bf H} ) = & \, \frac{1}{2} [ \bar J (U, {\bf H} ) + \bar J (U, -{\bf H} )], \quad \delta \bar J (U, {\bf H}) = \bar J (U, {\bf H} ) - \bar J (U, -{\bf H} ). 
\end{align}
To lowest order in $E_{T} /T$, the reciprocal part  is given by,
\begin{equation}\label{reciprocalcurrent}
\bar J_{\mathrm{rec}} (U, {\bf H} )= 2e\int^\infty_0 dN \partial_\epsilon n_F\big(\bar \epsilon(N, {\bf H}) \big) \sum_{k} |C_k(N, {\bf H})|^2  \bigg( \frac{ -k^2\lambda}{1 + (k\lambda)^2} \bigg).
\end{equation}
The nonreciprocal part of the current is given by,
%\begin{align}
%\frac{1}{2} \big( \bar J(U,{\bf H}) + \bar J(-U,{\bf H}) \big) =& 2e \int^\infty_0 dN \partial_\epsilon n_F\big(\bar \epsilon(N, {\bf H} ) \big) \sum_{k} |C_k|^2 \bigg( \frac{ik }{1 + (k\lambda)^2} \bigg) \nonumber \\
%&+\frac{e}{2\pi} \int^\infty_0 dN \partial^2_\epsilon n_F\big(\bar \epsilon(N, {\bf H} ) \big) \sum_{k_1,k_2,k_3} C_{k_1}(N) C_{k_2}(N) C_{k_3}(N)  \nonumber \\
%& \times \int^\pi_{-\pi} d\chi 
% \frac{ik_3 }{1 + (k_1 + k_2)^2 \lambda^2} e^{i(k_1 + k_2 + k_3)\chi}
%\end{align}
%Since $|C_k|^2 = |C_{-k}|^2$, the first term vanishes. To leading order in $1/T$, the nonreciprocal current is given by
\small
\begin{align}\label{nonreciprocalcurrent1}
\delta \bar J (U, {\bf H})  =& 2 e \int^\infty_0 dN \partial^2_\epsilon n_F\big(\bar \epsilon(N, {\bf H} ) \big) \sum_{k_1 , k_2 } C_{k_1}(N, {\bf H}) C_{k_2}(N, {\bf H}) C_{-k_1 - k_2}(N, {\bf H})  
%\nonumber \\
%& \times 
\bigg( \frac{-i(k_1 + k_2)^3 \lambda^2 }{1 + (k_1 + k_2)^2 \lambda^2} \bigg).
\end{align}
\normalsize
Note that the nonreciprocal current contains 2 derivatives of the Fermi function and is proportional to $(E_{T} /T)^2$, as opposed to the reciprocal current which is proportional to $E_{T} /T$.
%For a typical phase-dependence of the spectrum, the Fourier sums appearing in Eqs.~\eqref{reciprocalcurrent} and \eqref{nonreciprocalcurrent} are dominated by $k$ of order unity.

At small voltages $\lambda \ll 1$, the expressions for the reciprocal and nonreciprocal current can be simplified. 
To lowest order in $\lambda$, the reciprocal current is given by ,
\small
\begin{align}\label{reciprocalcurrent1}
\bar J_{\mathrm{rec}} (U, {\bf H} )&= (2e) \mathrm{sgn}(U) \int^\infty_0 dN \partial_\epsilon n_F\big(\bar \epsilon(N,{\bf H}) \big) \sum_{k} |C_k(N, {\bf H})|^2  \big(-k^2\lambda \big) \nonumber \\
&= 2e U \tau_{in} \int^\infty_0 dN \partial_\epsilon n_F\big(\bar \epsilon(N,{\bf H}) \big)  \frac{1}{2\pi} \int^\pi_{-\pi} d\chi \big( \partial_\chi \epsilon(N,\chi, {\bf H}) \big)^2 .
\end{align}
\normalsize
Similarly, the nonreciprocal current is given by
\begin{align}\label{nonreciprocalcurrent}
\delta \bar J (U, {\bf H})  
 =& 2e \int^\infty_0 dN \partial^2_\epsilon n_F\big(\bar \epsilon(N, {\bf H}) \big) \sum_{k_1 , k_2 } C_{k_1}(N,{\bf H}) C_{k_2}(N,{\bf H}) C_{-k_1 - k_2}(N,{\bf H})  \bigg( (i(k_1 + k_2)^3 \lambda^2) \bigg) \nonumber \\
 =& 2e(2eU\tau_{in})^2 \int^\infty_0 dN \partial^2_\epsilon n_F\big(\bar \epsilon(N,{\bf H}) \big) \frac{1}{2\pi} \int^\pi_{-\pi} d\chi  \big( \partial_\chi \epsilon(N, \chi, {\bf H}) \big)^3.
\end{align}
Changing variables back to $(\epsilon, \chi)$, we arrive at Eqs.~(12) and (13).
%Eqs.~\eqref{eq:G_l_g1} and \eqref{avgcurrent}
\color{black}
%\begin{align}\label{reciprocalcurrent2}
%\frac{1}{2}\big( \bar J(U,{\bf H}) - \bar J(-U,{\bf H}) \big) &= 2e\lambda \mathrm{sgn}(U) \int^\infty_0 d\epsilon \partial_\epsilon n_F\big(\epsilon \big)  \frac{1}{2\pi} \int^\pi_{-\pi} d\chi 
%\nu(\epsilon,\chi) V^2_\nu(\epsilon,\chi) \nonumber \\
%& = \langle g_1(\chi) \rangle (2eU\tau_{in})
%\end{align}
%\begin{align}\label{nonreciprocalcurrent2}
%\frac{1}{2}\big( \bar J(U,{\bf H}) + \bar J(-U,{\bf H}) \big) &= e\lambda^2 \int^\infty_0 d\epsilon \partial_\epsilon n_F\big(\epsilon \big)  \frac{1}{2\pi} \int^\pi_{-\pi} d\chi 
%\nu(\epsilon,\chi) V^3_\nu(\epsilon,\chi) \nonumber \\
%& = \langle g_2(\chi) \rangle (2eU\tau_{in})^2
%\end{align}
To estimate  the reciprocal and nonreciprocal current at relatively large voltages, $ 1/\tau_{in} \ll |2eU| \ll E_T $, we use the fact that for a typical phase-dependence of the quasiparticle spectrum, the sums over $k$ in Eqs.~\eqref{reciprocalcurrent} and \eqref{nonreciprocalcurrent1} are dominated by $k$ of order unity. With this in mind, we obtain the following estimates
\color{black}
\begin{align}\label{reciprocalLargeV}
\bar J_{\mathrm{rec}} (U, {\bf H} ) 
\sim& \frac{1}{U \tau_{in} } \int^\infty_0 dN \partial_\epsilon n_F\big(\bar \epsilon(N, {\bf H}) \big) \sum_{k} |C_k(N, {\bf H})|^2  ( -k^2 )
\sim \frac{\langle g_1(\chi, {\bf H}) \rangle }{eU\tau_{in}}, \\
\end{align}
\begin{align}\label{nonreciprocalLargeV}
\delta \bar J (U, {\bf H})  
 \sim& \frac{-ie}{(2eU \tau_{in} )^2} \int^\infty_0 dN \partial^2_\epsilon n_F\big(\bar \epsilon(N, {\bf H}) \big)  \sum_{k_1 , k_2 } C_{k_1}(N,{\bf H}) C_{k_2}(N,{\bf H}) C_{-k_1 - k_2}(N,{\bf H}) 
\sim \frac{\langle g_2(\chi, {\bf H}) \rangle }{(eU\tau_{in})^2}.
\end{align}
Combining Eqs.~\eqref{reciprocalLargeV} and \eqref{nonreciprocalLargeV} we get Eq.~(15)
%Eq.~\eqref{avgcurrent2} 
in the maintext.

The current as a function of the voltage is increasing at small voltages and decreasing at large voltages,  reaching a  maximum $J_{max} ({\bf H})$  in the intermediate regime. The location of  the maximum is found by setting to zero the derivative $\partial_\lambda \bar J$, 
\small
\begin{align}\label{derivativecurrent}
\partial_\lambda \bar J (U, {\bf H})   =&
-e \int^\infty_0 dN \bigg[   \mathrm{sgn}(U)\partial_\epsilon n_F\big(\bar \epsilon(N, {\bf H} ) \big) \sum_{k} |C_k(N, {\bf H})|^2  \bigg( \frac{2k^2 - 2k^4\lambda^2}{\big(1 + k^2 \lambda^2 \big)^2}\bigg) \nonumber \\
&+ \partial^2_\epsilon n_F\big(\bar \epsilon(N, {\bf H} ) \big) \sum_{k_1,k_2} C_{k_1}(N, {\bf H}) C_{k_2}(N, {\bf H}) C_{-k_1 - k_2}(N, {\bf H})  (k_1 + k_2)^2 \nonumber \\
& \times \bigg(\frac{\mathrm{sgn}(U)\big( 1 + \lambda^2(k_1 + k_2)^2\big) - 2\lambda(k_1 + k_2) \big(\mathrm{sgn}(U)\lambda (k_1 + k_2) + i \big)}{\big(1 + \lambda^2(k_1 + k_2)^2 \big)^2} \bigg)\bigg].
\end{align}
\normalsize
%The derivative of $\bar J$ changes sign as $\lambda$ is increased from $0$ to $+\infty$, as well as when it is decreased from $0$ to $-\infty$.
Since the sums are dominated by $k \sim 1$,
$\partial_\lambda \bar J $ vanishes at $\lambda \sim  1$. We can estimate the  reciprocal and nonreciprocal part of  $J_{max} ({\bf H})$ by substituting $\lambda =  1$ into Eqs.~\eqref{reciprocalcurrent1} and \eqref{nonreciprocalcurrent},
\color{black}
\begin{align}
J_{max}({\bf H})
 &\sim e\int^\infty_0 dN \partial_\epsilon n_F\big(\bar \epsilon(N, {\bf H} ) \big) \sum_k |C_k(N,{\bf H}) |^2
\bigg( \frac{-k^2}{1 + k^2} \bigg) \sim \langle g_1(\chi, {\bf H}) \rangle, \\
\delta J_{max}({\bf H})
&\sim -i e \int^\infty_0 dN \partial^2_\epsilon n_F\big(\bar \epsilon(N, {\bf H} ) \big) \sum_{k_1 , k_2 } C_{k_1}(N,{\bf H})  C_{k_2}(N,{\bf H})  C_{-k_1 - k_2}(N,{\bf H}) \sim \langle g_2(\chi, {\bf H}) \rangle .
\end{align}
{This reproduces Eq. (15) in the main text.}

\section{Appendix B: Derivation of Usadel's equations in diffusive junctions with spin-orbit coupling and a Zeeman field}
\label{sec:derivationUsadel}
\setcounter{equation}{0}
\renewcommand{\theequation}{B\arabic{equation}}

\newcommand{\ua}{{\uparrow}}
\newcommand{\da}{{\downarrow}}
\newcommand{\tin}{{\tau_{\mathrm{in}} }}
\newcommand{\tel}{{\tau_{\mathrm{el}} }}
\newcommand{\tso}{{\tau_{\mathrm{so}} }}
In this section we derive Usadel's equations  for the electron Green's functions in the diffusive approximation  in the presence of spin-orbit coupling and a Zeeman field.
We then we show that  in the special case of quasi-one dimensional geometry the I-U characteristics  of the junctions turn out to be reciprocal.
Thus, the non-reciprocity of the junctions is related to either a general character of {position-dependence} of the microscopic parameters describing the normal metal, such as $D({\bf r})$ and $\beta({\bf r})$, or  geometry of the normal region. 

%Although for some applications it is convenient to perform a unitary transformation of the semiclassical Green's functions to eliminate the linear in momentum spin-orbit coupling,~\cite{Aleiner-Falko,tokatly}, we choose not to do it.
%In contrast to Ref.~\cite{tokatly} the Usadel equation contains terms that are linear in spatial gradients and spin-orbit coupling,.
\subsection{Weak spin-orbit coupling}
{We will first consider the case of weak spin-orbit coupling, where $\beta p_F \ll \tau_{el}^{-1}$ }. The Hamiltonian describing a 2D SNS junction in the presence of weak spin-orbit coupling { and an in plane magnetic field} has the form  
\begin{equation}\label{Hamiltonian}
H =\left( \xi_p +\beta^{ij} \sigma^{i} p^j 
	 +  V_{imp}({\bf r}) \right)\tau_3 + g \mu_0 {\bf H} \cdot {\boldsymbol \sigma}  + \Delta({\bf r}) \tau_1 .
\end{equation}
Here $\xi_p = p^2 /2m  - E_F$, $\tau_{i}$ are the Pauli matrices in Nambu space, and $\Delta({\bf r})$ is the superconducting order parameter. We will focus on the case where the length of the normal metal region is much larger than the superconducting coherence length of the nodes $L \gg \xi$, and  $\Delta({\bf r})$ has the form
\begin{equation}\label{Delta}
\Delta ({\bf r}) = 
\begin{cases}
\Delta, & \quad x < - \frac{L}{2},\\
0, & \quad -\frac{L}{2} < x < \frac{L}{2},\\
\Delta e^{-i\chi}, & \quad x> \frac{L}{2}.
\end{cases} 
\end{equation}
We start with the Eilenberger equation  \cite{eilenberger} corresponding to the Hamiltonian in Eq.~\eqref{Hamiltonian}
\color{black}
\begin{align}
\label{eq:Eilenberger}
	\left[ \epsilon\underline\tau_3 -  \Delta({\bf r}) \tau_1 + A_0 \underline\tau_3, \underline g \right] + i v_F \hat p^k \tilde\nabla_{\bf r}^k  \underline g - {\frac{i}{2m}} \left\{ A_k, \nabla_{\bf r}^k  \underline g \right\} = & \,  \frac{i}{2 \tau_{el}} \left[  \underline g_{0}, \underline g \right], &&
\underline g^2 =   1.
\end{align}
Here $\underline g({\bf r}, \epsilon, {\bf \hat p}) $ is the quasi-classical Green's function, which is a matrix in Nambu-spin space, {  $\{ \cdot, \cdot \}$  and $[\cdot, \cdot]$ denote the anti-commutator and commutator  respectively,  $ \hat p$ is the unit vector pointing in the direction of $\vec p$}, and $g_{0}$ is the Green's function averaged over the direction of $ \hat p$. Finally, $ {\bf h} = g \mu_0 {\bf H}$ is the Zeeman energy, 
  \begin{align}
A_0 = & \,  - h^\alpha \sigma^\alpha, \qquad A_k =  {-m \beta^{\alpha k} \sigma^\alpha} ,
\end{align}
and  $\tilde\nabla_{\bf r}^k   = \nabla^k_{\bf r} - i[A_k, \cdot] $. 
 
 We note that  the third Eq.~\eqref{eq:Eilenberger} differs from the Eilenberger equation used in Ref.~\cite{tokatly}. The reason for this is because in Ref.~\cite{tokatly} the  author chooses to perform a unitary transformation of the semiclassical Green's functions which eliminates terms which are linear in momentum and spin-orbit coupling, we choose not to do this.
 The density of sates in the normal region can be written
  in terms of the Green's function {in the form,} 
\begin{equation}\label{DOS1}
\nu(\epsilon) = \frac{\nu_N}{4 L_1 L} \int d {\bf r} \Re \big[ \mathrm{Tr}_{\tau, \sigma} \big( \tau_3 \underline g_0(\epsilon, {\bf r}) \big) \big].
\end{equation}
Here $\nu_N$ is the total density of states of the normal metal part of the junction in the absence of the superconductors, $\mathrm{Tr}_{\tau, \sigma} $ {denotes} the trace over Nambu-spin space, and the integral is taken over the {area} of the normal region.

In the diffusive regime where the Green's functions are nearly isotropic, we can expand the Green's function into its zeroth and first {angular harmonics},  
%\begin{subequations}
\begin{align}
\label{normalization}
	\underline g = &  \underline g_0 + \underline g_1^k \hat p^k ,\quad 
	|\underline{\bm g}_1| \ll |\underline g_0|.
\end{align}
%\end{subequations}
With these assumptions, the normalization conditions are then 
%\begin{subequations}
\begin{align}\label{g0g1}
	\underline g_0^2 &= 1,\quad 
	\left\{ \underline g_0, \underline g_1^k \right\} = 0.
\end{align}
%\end{subequations}
Substituting Eq.~\eqref{normalization} into Eq.~\eqref{eq:Eilenberger}, and using Eq.~\eqref{g0g1} we get the equations for $g_{0}$ and $g_{1}^{k}$
\begin{subequations}
\begin{align}\label{g0}
	\left[ \epsilon\underline\tau_3 -  \Delta({\bf r}) \tau_1 + A_0 \underline\tau_3, \underline g_0 \right] + \frac{i v_F}{2} \tilde\nabla_{\bf r}^k  \underline g_1^k  - {\frac{i}{2m}  } \left\{ A_k, \nabla_{\bf r}^k  \underline g_0 \right\} =& \,  0, \\ \label{g1}
	 v_F \tilde\nabla_{\bf r}^k  \underline g_0 -  \frac{1}{2\tel} \left[ g_0, \underline g_1^k \right] = & \, 0.
\end{align}
\end{subequations}
Multiplying  both sides of Eq.~\eqref{g1} by $\underline g_0$ and using  Eq.~\eqref{g0g1}, we get
\begin{align}
\label{eq:1stMomentSolution}
	\underline g_1^k &= - v_F \tel \underline g_0 \tilde\nabla_{\bf r}^k  \underline g_0.
\end{align}
Substituting Eq.~\eqref{eq:1stMomentSolution} back into Eq.~\eqref{g0}, we arrive at Usadel's equation,
\begin{align}
\label{eq:UsadelNambuSpin}
	\left[\epsilon\underline\tau_3 - \Delta({\bf r}) \tau_1 + A_0 \underline\tau_3, \underline g_0 \right] - i  \tilde\nabla_{\bf r}^k  \left( D({\bf r}) \underline g_0 \tilde\nabla_{\bf r}^k  \underline g_0 \right) - {\frac{i}{2m}} \left\{ A_{k}({\bf r}), \nabla_{\bf r}^k  \underline g_0 \right\} &= 0.
\end{align}
To account for long range variations of the microscopic parameters, we allow $ D({\bf r})$  and $A_{k}({\bf r})$ to depend on position.
We note that Eq.~\eqref{eq:UsadelNambuSpin} is valid provided that $ D({\bf r})$  and $A_{k}({\bf r})$ change slowly on the scale of the { spin relaxation length}. 
  
 %Note that the effects of spin-momentum locking are incorporated into the covariant derivatives in the second term, whereas the spin-charge transfer is incorporated into the third term.
Equation \eqref{eq:UsadelNambuSpin} is written in Nambu-spin space. If $h \ll \tau_{so}^{-1}$ , it is useful to decompose to the Green's functions into a singlet and triplet components, 
\begin{align}
\label{gs_gt}
	\underline g_0 = \underline g_s + \underline g_t^\alpha \sigma^\alpha, \quad
	|\underline g_t^\alpha| \ll |\underline g_s|.
\end{align}
Here $\underline g_s$, $\underline g_t^\alpha$ are matrices in Nambu space.
In this approximation, the normalization conditions Eq.~\ref{g0g1}  yield
%\begin{subequations}
\begin{align}
\label{eq:SpinNormalization}
	\underline g_s^2 &= 1, \quad 
	\left\{\underline g_s, \underline g_t^\alpha \right\} =0.
\end{align}
%\end{subequations}
Substituting Eq.~\eqref{gs_gt} into \eqref{eq:UsadelNambuSpin}  and using  Eq.~\eqref{eq:SpinNormalization}, we get 
\begin{subequations}
\begin{align}
\label{eq:SingletU1}
\left[\epsilon\underline\tau_3 - \Delta({\bf r}) \tau_1, \underline g_s \right] - i \nabla_{\bf r}^k  \left(D({\bf r})\underline g_s \nabla_{\bf r}^k  \underline g_s \right) - h^\alpha \left[ \underline\tau_3, \underline g_t^\alpha \right] + i \beta^{\alpha k}({\bf r}) \nabla_{\bf r}^k  \underline g_t^\alpha & \, =0, \\
\label{eq:TripletU1}
\frac{i}{\tau_{so}} \underline g_s \underline g_t^\alpha + \left[ \epsilon\underline\tau_3 -  \Delta({\bf r}) \tau_1, \underline g_t^i \right] 
- i  \nabla_{\bf r}^k  \left( D({\bf r})  \underline g_s \nabla_{\bf r}^k  \underline g_t^i + \underline g_t^i \nabla_{\bf r}^k  (D({\bf r}) \underline g_s) \right) & \nonumber  \\
+2im D({\bf r}) \epsilon^{i\alpha\beta} \beta^{\alpha k}({\bf r})  \left( \underline g_s \nabla_{\bf r}^k  \underline g_t^\beta + \underline g_t^\beta \nabla_{\bf r}^k  \underline g_s + \nabla_{\bf r}^k  \left(\underline g_s \underline g_t^\beta\right) \right)  
& \, = h^i \left[\underline\tau_3, \underline g_s \right] -i\beta^{i k}({\bf r})   \nabla_{\bf r}^k  \underline g_s.
\end{align}
\end{subequations}
%where $\tau_{so}^{-1} = 4 p_F^2 \beta^2 \tau_{el}$.
Note that we have dropped terms $\mathcal O \left( \underline g_t^2 \right) $, and that only ordinary derivatives are left in the equations. 
%The spin-momentum locking manifests itself in the linear and quadratic in $\beta$ terms on the LHS of Eq.~\eqref{eq:TripletU1}, meaning that spin-momentum locking cannot generate a triplet component from a singlet component and vice versa. Spin-momentum locking instead serves as a source of relaxation for $\underline g_t^\alpha$.
The terms on the RHS of Eq.~\eqref{eq:TripletU1} act as source terms for the triplet Green's functions. Thus, the triplet Green's function is generated either by the Zeeman field or the linear in gradient terms {arising from} spin orbit coupling. 

Since we are interested in solutions for $\epsilon \sim E_T$ and $\Delta({\bf r}) = 0$ in the normal metal, the second term on the LHS of Eqs.~\eqref{eq:SingletU1} is on the order of $E_T$.
The typical length scale on which the Green's functions change is on the order of $L$ (see Ref.~\cite{ZhouSpivak}), so the gradients in Eqs.~\eqref{eq:SingletU1} and \eqref{eq:TripletU1} are of order $\frac{1}{L}$. 
With this in mind, we note that the third term on the LHS of Eq.~\eqref{eq:TripletU1} is also on the order of $E_T$.  In the case  $E_T \ll \tau_{so}^{-1}$ the  first term in Eq.~\eqref{eq:TripletU1} is the largest term on the LHS,  and we get
\color{black}
%\begin{align}
%\label{eq:TripletU2}
%	\left[\epsilon\underline\tau_3, \underline g_t^\alpha \right] + \frac{i}{\tau_{so}} \underline g_s \underline g_t^\alpha & = h^\alpha \left[ \underline\tau_3, \underline g_s \right] - i \beta^{\alpha k} \nabla_{\bf r}^k  \underline g_s.
%\end{align}
%We substitute $ g_s = \tau_3$ on the LHS, which is acceptable to leading order in $\tel$. Using the normalization conditions Eqs.~\eqref{eq:SpinNormalization} we have
\begin{align}
\label{eq:TripletUFinal}
	\underline g_t^\alpha &= -i\tau_{so} \left( \underline g_s \left[ h^\alpha\underline\tau_3, \underline g_s \right] -i \beta^{\alpha k} ({\bf r}) \underline g_s \nabla_{\bf r}^k  \underline g_s \right).
\end{align}
Finally, substituting Eq.~\eqref{eq:TripletUFinal} into Eq.~\eqref{eq:SingletU1}, we get the Usadel equation for $\underline g_s$
\begin{align}
\label{eq:SingletU2}
	\left[\epsilon\underline\tau_3, \underline g_s \right] 
- i \nabla_{\bf r}^k  \left(D({\bf r}) \underline g_s \nabla_{\bf r}^k  \underline g_s \right) 
+ i\tau_{so} h^2 \left[ \underline \tau_3, \underline g_s [ \underline \tau_3, \underline g_s ] \right]&  \nonumber \\
+ \tau_{so}  \beta^{\alpha k}({\bf r})  h^\alpha \left( \left[ \underline\tau_3, \underline g_s \nabla_{\bf r}^k  \underline g_s \right] + \nabla_{\bf r}^k  \left(\underline g_s \left[ \underline\tau_3, \underline g_s\right] \right) \right) & = 0.
\end{align}
 In the subsequent equations we only keep terms which are linear in $h$, as we will be interested only in the part of the density of states which is linear in ${\bf H}$. 
\color{black}

Let us first consider junctions with a 1D geometry, where $D({\bf r})$ as well as $\beta ({\bf r})$ depend only on the $x$ coordinate.  In this case, Eq.~\eqref{eq:SingletU2} and Eq.~\eqref{DOS1} are given by
\begin{align}
\label{eq:SingletU3}
	\left[\epsilon\tau_3,  g_s \right] - i \partial_x  \left( D(x) g_s \partial_x  g_s \right)  +\tau_{so} \beta^{\alpha x}(x  h^\alpha \left( \left[ \tau_3, g_s \partial_x  g_s \right] + \partial_x   \left( g_s \left[ \tau_3, g_s\right] \right) \right) = 0, & \\
\nu(\epsilon) = \frac{\nu_N}{2L} \int^{L/2}_{-L/2} d x \Re \big[ \mathrm{Tr}_{\tau}\big( \tau_3 \underline g_s(\epsilon, x) \big) \big] &,
\end{align}
\color{black}
where and $\mathrm{Tr}_{\tau}$ is the trace over Nambu space.

It is convenient to parameterize the Green's functions in the following form, 
\begin{align}
	g_s (\epsilon,x) = \left(\begin{array}{cc}
		\cos\theta(\epsilon,x) & \sin\theta(\epsilon,x) e^{i \tilde \chi(\epsilon,x)}\\
		\sin\theta(\epsilon,x) e^{-i \tilde \chi(\epsilon,x)} & -\cos\theta(\epsilon,x)
	\end{array}\right),
\end{align}
where $\theta(\epsilon,x)$ and $\tilde \chi (\epsilon,x)$ are complex variables. In this {parametrization} the Usadel equation reduces to the following two equations,
\begin{align}
\label{eq:ParamEq1}
	\partial_x  \left[  \big(D(x) \partial_x  \tilde \chi + 2\tau_{so} \beta^{\alpha x} (x)  h^\alpha \big) \sin^2\theta  \right] & = 0,  \\
\label{eq:ParamEq2}
 \frac{1}{2} \partial_x \bigg(D(x) \partial_x \theta \bigg)   + i\epsilon\sin\theta 
- \frac{D(x)}{4}\sin 2\theta \bigg(\partial_x \tilde \chi  + \frac{2\tau_{so} \beta^{\alpha x}( x)  h^\alpha}{D(x)} \bigg)^2
 & = 0,
\end{align}
and the expression for the density of states becomes
\begin{align}\label{DOS4}
	\nu(\epsilon,\chi) &=  \frac{\nu_N}{L} \int^{L/2}_{-L/2}  d {x}  \Re \cos\theta(\epsilon, x ) .
\end{align}
In the case of perfectly transmitting NS interfaces, the boundary conditions for Eqs.~\eqref{eq:ParamEq1},   \eqref{eq:ParamEq2} are given by
%\begin{subequations}
\begin{align}\label{BC}
	\theta (\epsilon, \pm L/2) &= \frac{\pi}{2}, \quad 
	\tilde \chi(\epsilon, \pm L/2) =  \pm \frac{\chi}{2}.
\end{align}
%It is worth mentioning that Eqs.~\eqref{eq:ParamEq1} and \eqref{eq:ParamEq2} have a first integral, and can be reduced to a set of first order differential equations. The exact solutions of the corresponding first order equations can then be expressed in terms of elliptical functions. However, for our purposes it is more convenient to work with Usadel's equations in the form of Eq.~\eqref{eq:ParamEq1}-\eqref{eq:ParamEq2} , and we will not focus on this procedure of obtaining exact solutions.
We can further simplify these equations by making the change of variables $(\theta, \tilde \chi) \rightarrow (\theta, \hat\chi)$, where the shifted phase $\hat{\chi}$ is given by
\begin{equation}
\hat \chi = \tilde \chi + \int_{-L/2}^{x} dq \frac{2\tau_{so} \beta^{\alpha x} (q) h^\alpha }{D(q)}. 
\end{equation}
In the new variables Eqs.~ \eqref{eq:ParamEq1}, \eqref{eq:ParamEq2}, and \eqref{BC} become
\begin{align}
\label{eq:ParamEq3}
	\partial_x  \left[ D(x) \partial_x \hat \chi \sin^2\theta  \right] & = 0,  \\
\label{eq:ParamEq4}
 \frac{1}{2} \partial_x\bigg( D(x) \partial_x \theta \bigg)   + i\epsilon\sin\theta 
- \frac{D(x)}{4}\sin 2\theta \big(\partial_x \hat \chi\big)^2
%+ \frac{D}{4}\sin 2\theta  \bigg(\frac{4\tau_{so} \beta^{\alpha x} h^\alpha}{D} \bigg)^2
 & = 0,\\
\label{BC2}
	\theta (\epsilon, \pm L/2) = \frac{\pi}{2}, \qquad 
	\tilde \chi(\epsilon, \pm L/2) =  \pm \bigg(\frac{\chi + \phi}{2} \bigg) & .
\end{align}
We note  that Eqs.~\eqref{eq:ParamEq3}--\eqref{BC2} are identical to Usadel's equations and their boundary conditions  at $ h = 0$, with a phase difference of $\chi + \phi$ across the junction. Thus, the solution for $\theta$ has the form
\begin{align}\label{shift1}
\theta(\epsilon,x,\chi) = \theta_0(\epsilon, x ,\chi + \phi)
\end{align}
 Here $\theta_0(\epsilon, x ,\chi )$ is the solution for zero magnetic field and phase difference $\chi$, and $\phi$ is given by Eq.~(27). Thus, we see that the phase dependence of  the density of states has the form of Eq. (10).
As discussed in the main text, in this case the I-U characteristics are reciprocal.

We would like to stress however that  this result {arose from the 1D character of the idealized model of the the  SNS junction. In the general situation, where the parameters $D({\bf r})$ and $\beta({\bf r})$ are functions of $x$ and $y$, or the shape on the N-region is non-rectangular,  the density of states $\nu(\chi)$ can not be expressed in the form of Eq.~(25),  and the I-U characteristics are nonreciprocal.}

Consider for example a simple model where the diffusion coefficient is a function of $y$, a coordinate parallel to the SN interface,
%\color{blue}
\begin{equation}
D(x,y) = \begin{cases}
D  + \frac{\delta D}{2} ,& 0 < y < L_1/2 ,\\
D - \frac{\delta D}{2}, & -L_1/2< y <0 ,
\end{cases}
\end{equation}
where $\delta D \ll D$. 
If $L_{1}\gg L$ then the local density of states in the regions with $y>0$ and $y< 0$ can be approximated ${\bf H} =0 $ solutions shifted by phases  $\phi_{\pm} =  \frac{2\tau_{so} \beta^{\alpha x} h^\alpha L}{D \pm \delta D/2 }$ respectively.
Then the total density of states can be written as  
\begin{equation}\label{DOS3}
\nu(\epsilon, \chi, D) \approx \frac{1}{2} \bigg( \nu_0(\epsilon, \chi + \phi_+, D + \delta D/2) + \nu_0(\epsilon, \chi + \phi_-, D - \delta D/2) \bigg),
\end{equation}
where $\nu_0(\epsilon, \chi, D )$ is density of states of a junction with diffusion coefficient $D$ and dimensions $L_1 \times L$,  at ${\bf H}= 0$. In this case the total density of states cannot be expressed in the form of equation of Eq.~(25), and the I-U characteristics are non-reciprocal. Substituting Eq.~\eqref{DOS3} into Eq.~(17b) in the main text and keeping terms linear in ${\bf H}$, we get the following estimate for $\langle g_2 \rangle$,
\begin{align}\label{g2estimate}
\langle g_2(\chi, {\bf H}) \rangle  &\approx 
 \bigg( \frac{\tau_{so} \beta^{\alpha x} (g \mu_0 {\bf H}^\alpha)}{L E_T} \bigg)  \bigg(\frac{\delta D}{D} \bigg)^2 \frac{E_T^2}{T^2} \langle g_1(\chi,{\bf H}) \rangle.
\end{align}
Similarly, we substitute Eq.~\eqref{DOS3} into Eq.~(22) to obtain the following estimate for $\delta A$,
\begin{align}\label{deltaAestimate}
\delta  A  & \approx \frac{\sqrt{J_c(0,T)} }{e\tau_{in} \langle g_1 \rangle}
\frac{\beta ( \tau_{so} g \mu_0 H) }{L E_T }  \bigg( \frac{\delta D}{D} \bigg)^2 .
\end{align}
Here we have used the fact that  both the critical current and the phase $\chi_m({\bf H})$, at which the supercurrent reaches its maximum, are functions of $\nu(\epsilon,\chi,D)$.

In the more general case where $L_{1}\gtrsim L$, the scale of spacial fluctuations of $D({\bf r})$ and $\beta({\bf r})$ are of order $L$,  and $\delta D \lesssim D$, we expect these estimates to be accurate to order unity.

\subsection{Superconductor-topological insulator-superconductor (S-TI-S) junction}
{In this section we consider a model in which the normal region of the junction is comprised of a conducting surface of a topological insulator described by
the Hamiltonian
\begin{align}\label{hamiltonianSSO}
	\mathcal H(k) = \left( \beta^{\alpha k} k^i \sigma^\alpha - \mu + V_{dis}\right) \tau_3 + h^\alpha \sigma^\alpha + \Delta \tau_1.
\end{align}
This model can be viewed a limiting case of strong spin-orbit coupling in which the conduction band has a definite helicity. }

The derivation of the Usadel equation in S-TI-S junction was done by Ref.~\cite{Loss}; here we will sketch the main steps.
The Eilenberger equation for the Hamiltonian in Eq.~\eqref{hamiltonianSSO} is given by,
\begin{align}
\label{eq:TIEilenberger}
	\left[ \epsilon\underline\tau_3 -  \Delta({\bf r}) \tau_1 - h^\alpha \sigma^\alpha \underline\tau_3, \underline g(\bm n, \epsilon, {\bf r}) \right] + p_F \beta^{\alpha k}\hat p^k \left[ \sigma^\alpha \underline g({\bf \hat p},\epsilon,\bm r) \right] 
+ \frac{i}{2} \beta^{\alpha k} \left\{ \sigma^\alpha, \nabla_{\bf r}^k \underline g({\bf \hat p},\epsilon,\bm R) \right\} = & \, - \frac{i}{2\tel} \left[ \langle \underline g \rangle, \underline g({\bf \hat p}, \epsilon,\bm R) \right], 
\end{align}
 In the regime of strong spin orbit coupling where the system has complete spin-momentum locking, the Green's function has the following helical spin structure, 
\begin{align}
\label{eq:gprime}
	\underline g = \frac{1}{2} \underline g' \left(1 + \hat \beta^{\alpha k} \hat p^k \sigma^\alpha \right),
\end{align}
where $\underline g'$ is a matrix only in Nambu space. 
We then insert Eq.~\eqref{eq:gprime} into Eq.~\eqref{eq:TIEilenberger},
\begin{align}
	\left[ \epsilon\underline\tau_3, -  \Delta({\bf r}) \tau_1- h^\alpha\sigma^\alpha, \frac{1}{2}\underline g' \left(1 + \hat \beta^{\gamma k }{ \sigma^ \gamma} \hat p^k \right) \right] 
+ip_F \epsilon^{\alpha\beta\gamma} \beta^{\alpha k} \hat\beta^{\beta i} \hat p^k \hat p^i g' \sigma^\gamma  &
	\nonumber \\
 + \frac{i \beta^{\alpha k}}{2}  \nabla_{\bf r}^k g' \left( \sigma^\alpha + \hat \beta^{\alpha i} \hat p^i \right)
\label{eq:TIEilenberger2}
+ \frac{i}{2\tel} \left[ \left\langle \frac{1}{2}\underline g' \left(1 + \hat \beta^{\alpha k} \hat p^k \right)  \right\rangle, \frac{1}{2}\underline g' \left(1 + \hat \beta^{\alpha k} \hat p^k \right) \right]
	& = 0.
\end{align}

In the diffusive limit where $v_F \tau_{el} \ll L$,
% we follow the procedure outlined in Eqs.~\eqref{normalization} - \eqref{eq:UsadelNambuSpin} for the weak spin orbit case, 
we expand the Green's function into its zeroth and first moment $\underline g' = \underline g_0 + \underline g_1^k \hat p^k$, insert the expansion into Eq.~\eqref{eq:TIEilenberger2}, project the equation onto its zeroth and first angular harmonics, and obtain two equations for $\underline g_{0}, \underline g_1$. After taking the trace of the resulting equations over spin indices, we arrive at the following equations,
\begin{align}\label{g0sso}
	\left[ \epsilon\underline\tau_3 -  \Delta({\bf r}) \tau_1, \underline g_0 \right] + i \frac{\beta}{2} \hat \nabla_{\bf r}^i \underline g_1^i = 0,
\end{align}
\begin{align}\label{g1sso}
	\underline g_1^i &= - 2 \beta \tel \underline g_0 \hat\nabla_{\bf r}^i \underline g_0,
\end{align}
where we have defined,
\begin{align}
	\hat\nabla_{\bf r}^i \cdot = \nabla_{\bf r}^i \cdot + i \frac{\hat\beta^{\alpha i} h^\alpha}{\beta} \left[ \underline\tau_3, \cdot \right].
\end{align}
%%%%%%%%%%%%%%%%%%%%%%%%%%%%%%%%%%%%%%%%%%%%%%%%%%%%%%%%%%%%%%
Substituting Eq.~\eqref{g1sso} in \eqref{g0sso}, we have Usadel's equation for the strong spin orbit case,
%\begin{align}
%	\left[\epsilon \underline\tau_3 -  \Delta({\bf r}) \tau_1 , \underline g_0 \right] - i D \hat\nabla_{\bf r}^i \left( \underline g_0 \hat\nabla_{\bf r}^i \underline g_0 \right) = 0,
%\end{align}
%where ${\color{blue} D = \beta^2\tel }$. Expanding out the covariant derivatives, we have
\begin{align}\label{singletU22}
	\left[ \epsilon\underline\tau_3 - \Delta({\bf r}) \tau_1, \underline g_0 \right] 
- i  \nabla_{\bf r}^i \left(  D({\bf r})\underline g_0 \nabla_{\bf r}^i \underline g_0 \right) 
+ \tau_{el}  \beta^{\alpha i}({\bf r})  h^\alpha \left( \nabla_{\bf r}^i \left( \underline g_0 \left[ \underline\tau_3, \underline g_0 \right] \right)   +\left[ \underline\tau_3, \underline g_0 \nabla_{\bf r}^i \underline g_0 \right] \right)  
+ i  \tau_{el} h^2 \left[ \underline\tau_3, \underline g_0 \left[ \underline\tau_3, \underline g_0 \right] \right] =& \,  0,
\end{align}
which is the same as Eq.~\eqref{eq:SingletU2},  the equation for $g_0$ in the case of weak spin-orbit coupling, with the replacement of $ \tau_{so}$ with $\tau_{el}$ .  
This can be understood by the fact that in the case where spin-momentum locking is strong, spin-relaxation is limited only by the rate of elastic scattering. 
This suggests that the form of Eq.~\eqref{singletU22} is universal within the diffusive regime, where the parameters in the equation depend on the strength of the spin-orbit coupling.
As a result, we can apply the estimates for $\langle g_2 \rangle$ and $\delta A$ given in Eqs.~\eqref{g2estimate} and \eqref{deltaAestimate} to the case of a S-TI-S junction with the appropriate substitution of $\tau_{so} = \tel$.
\color{black}
\end{widetext}

\end{document}